\documentclass{jstatphys}
\usepackage[reqno,sumlimits]{amsmath}
\usepackage{amsfonts}
\usepackage{amssymb}
\usepackage{bbm}
\usepackage{epsfig}

\def\be{\begin{equation}}
\def\ee{\end{equation}}
\def\bc{\begin{center}}
\def\ec{\end{center}}

\begin{document}

 

\title{Crossing the coexistence line at constant magnetization}

\author{Michel~Pleimling%
\footnote{Institut f\"ur Theoretische Physik 1, Friedrich-Alexander-Universit\"at 
Erlangen-N\"urnberg, Staudtstra{\ss}e 7, D -- 91058 Erlangen, Germany.}
\footnote{ Laboratoire de Physique des Mat\'eriaux,
Laboratoire associ\'e au CNRS UMR 7556,
Universit\'e Henri Poincar\'e Nancy I, B.P. 239, 
F -- 54506 Vand{\oe}uvre l\`es Nancy Cedex, France}
and Alfred~H\"{u}ller%
\addtocounter{footnote}{-2}
\footnotemark
}

\runningauthor{Pleimling and H\"{u}ller}

\date{Version of \today}


 
\begin{abstract}
Using Monte Carlo histogram methods, the microcanonical caloric curve is computed
for the Ising model in two and three dimensions with fixed magnetization. Whereas
the signatures of a possible first order phase transition are clearly visible for large
systems, intriguing finite size effects are revealed for smaller system sizes.
The behaviour of the caloric curve is studied in a systematic way. Furthermore,
results for the thermal stability of three-dimensional droplets of minority spins inside the
two-phase region are presented. The effect of the percolation transition  on the stability
of these droplets is discussed.
\end{abstract}

\section{Introduction}
The celebrated Ising model has in the past been studied extensively, mainly
because of its conceptual simplicity in combination with non-trivial features 
and because of the fact that,
in two dimensions, exact results can be obtained\cite{Mcc73}. 
Besides being the best studied model in the
theory of critical phenomena, it is for example also discussed in connection
with magnetism, surface physics, or even nuclear physics
(see Ref.\ \cite{Ric00} for a recent overview on the use of classical
microscopic models in nuclear physics).

In the present work we want to focus on the behavior of
the two- and three-dimensional nearest neighbour Ising models
inside the coexistence region. In the temperature-magnetization space
this two-phase region is bordered by the line of spontaneous magnetization,
the coexistence line.
Investigations inside the two-phase region mainly deal with 
non-equilibrium dynamics\cite{Bin87,Bra94}
(analyzing, for example,  the time evolution when
the external field is switched or when the system is quenched from 
high temperatures) or with the equilibrium shape of droplets of
minority spins\cite{Rot84,Dob92,Bod00}.
Using Monte Carlo techniques we study the Ising model
in the temperature-magnetization  and in the energy-magnetization space
when the magnetization density $m$ is fixed
at some value $m_0$. 
This approach has the advantage that data both below and above the
coexistence line can be gathered without having to deal with the 
complications of an 
external field changing sign at the coexistence line.

Recently, properties of a surface with a constant coverage of adatoms
have been described within a model equivalent to the 
two-dimensional Ising model with fixed magnetization (IMFM)\cite{Mul99}. 
The transition between a cluster
of monoatomic height at low temperatures and a gas of adatoms at higher
temperatures, observed in that study, has led to a subsequent work on the two-dimensional IMFM where the
finite-size dependences of the stability of minority clusters
in small systems were elucidated\cite{Ple00}. 
These system are small in comparison with the asymptotically large systems
where rigorous results are available\cite{Dob92,Bod00,Dob94,Iof98}.
In particular, numerical simulations of systems of
finite size showed that the macroscopic droplets of minority spins (i.e. clusters whose
radii scale with the system size) which 
disappear at the coexistence line in the thermodynamic limit, do that
with the signatures of a possible phase transition as e.g.
pronounced maxima in cluster and thermal properties. 
In three dimensions, the IMFM model was
considered in relation to 
nuclear matter fragmentation\cite{Car98}. 
Here, based on simulations of moderate sized
systems, the cluster transition was discussed in terms of a 
continuous transition taking place at the coexistence line.
To be complete, we should also mention a recent study of the 
three-dimensional IMFM
at the critical temperature\cite{Blo00}.

Based on general considerations on the topology
of the entropy of finite systems\cite{Gro00}, one would expect 
to observe the finite-size signatures
of a discontinuous transition when crossing this line
at fixed magnetization, but recent numerical studies of finite systems in two\cite{Ple00}
and three\cite{Car98} dimensions seemed to be more compatibel with a continuous
transition. Our main subject will be
the microcanonical caloric curve as obtained
from Monte Carlo simulations of the IMFM in two and three dimensions
(analysing a vast range of different lattice sizes). Indeed, a thorough
analysis of the caloric curve is expected to provide new insights.
Furthermore, as a complement to Ref.\cite{Ple00} , we will also 
discuss the influence of finite-size effects on the thermal stability
of macroscopic droplets of minority spins in three dimensions.

The paper is organized as follows. In the next Section our simulation
techniques are introduced. Especially, we want to point out 
how histogram methods can be used for the computation of the
caloric curve. In Section 3, microcanonical caloric curves of the IMFM in two 
and three dimensions are discussed in detail,
whereas the thermal stabilty of three-dimensional extensive droplets on finite 
lattices is investigated in
Section 4. Finally, Section 5 gives our conclusions.

\section{Method}
Monte Carlo (MC) calculations have been developed into a universally 
applicable tool for problems which defy an exact solution by  
analytical methods. In statistical mechanics many of these extremely 
difficult problems occur in the vicinity of phase transitions. There the 
MC method serves the purpose of calculating the mean values of 
physically interesting quantities as e.g. the energy density or the 
magnetization density of the system. The dissipation fluctuation 
theorem relates the response functions to the mean values of the 
spontaneous fluctuations of the corresponding densities.  For the 
examples mentioned above the fluctuations yield the specific heat 
and the magnetic susceptibility, respectively.

The main purpose of MC calculations is thus the determination of 
mean values of densities and of their fluctuations.

Most calculations are performed for the experimentally relevant 
situation where the system of interest is in contact with a thermostat. 
There the microstates $\mu$  of the system appear according to the 
canonical distribution
\begin{equation} \label{Gl:1}
p_\mu = \frac{\exp \left( - \beta E_\mu \right)}{Z}
\end{equation}
where $ E_\mu$ is the value of the energy in the microstate $\mu$,
$\beta$ is the inverse temperature of the thermostat and $Z$ the partition 
function
\begin{equation} \label{Gl:2}
Z = \sum\limits_\mu \, \exp \left( - \beta E_\mu \right).
\end{equation}
In conventional MC calculations an approximation for the mean value 
of the energy
\begin{equation} \label{Gl:3}
\left< E \right> = \sum\limits_\mu \, p_\mu E_\mu
\end{equation}
is directly obtained in the course of the MC run by sampling the 
energy values $E_t$ at $\nu$ sufficiently distant instants $t$ 
of the calculation
\begin{equation} \label{Gl:4}
\left< E \right> \cong  \frac{1}{\nu} \sum\limits_t E_t.
\end{equation}
The mean value of the fluctuations are obtained by summing the 
equivalent expression for $\left< \left( E - \left< E \right> \right)^2
\right>$.

In recent years, however, the direct summation has frequently been 
replaced by the histogram method\cite{Fer88}. There one tries to find an 
approximation for the degeneracy $g(E_i)$ of the energy levels $E_i$
of the system
\begin{equation} \label{Gl:5}
 g(E_i) = \sum\limits_\mu \delta_{E_i,E_\mu}.
\end{equation}
To this end the energy values $E_t$ at the instants $t$  are collected into 
a histogram: 
\begin{equation} \label{Gl:6}
 g(E_i) \, \exp \left( - \beta E_i \right) = K \, \sum\limits_t \,
\delta_{E_i,E_t}.
\end{equation}
Eq.\ (\ref{Gl:6}) can be solved for $g(E_i)$ in the energy range where the 
statistical errors  on the right hand side are sufficiently small. 
$K$ is an undetermined constant.

Nothing is lost when a histogram is stored instead of the mean 
values. On the contrary there are several advantages: (1) Once 
$g(E_i)$ has been established, the canonical mean value of any 
function $X(E)$ of the energy can easily be calculated afterwards:
\begin{equation} \label{Gl:7}
\left< X \right>_{\beta '} = \sum\limits_i \, g(E_i) \, \exp \left(
- \beta ' E_i \right) \, X(E_i) / Q_c
\end{equation}
with
\begin{equation} \label{Gl:8}
Q_c = \sum\limits_i \, g(E_i) \, \exp \left( - \beta ' E_i \right).
\end{equation}
(2) As indicated by the prime in the formula (\ref{Gl:7}), $\beta '$
may deviate by a small amount from the value of $\beta$ which has been used in the 
simulation where $g(E_i)$ was established. This allows the determination of 
the temperature dependence of $\left< X \right>$ in a small range around 
$T = 1/\beta$. (3) With an algorithm that finds the best values for the
undetermined constants $K$, histograms of MC runs at several temperatures 
may be combined to yield a single broad histogam which may then be used to
find the temperature dependence of $\left< X \right>$ 
over the desired range\cite{Fer88}. 

Here we shall exploit yet another advantage of the histogram 
method. (4) From the entropy $S(E)= \ln g(E)$ and the
microcanonical definition
\begin{equation} \label{Gl:9}
\beta^*(E) = \mbox{d}S(E)/\mbox{d}E
\end{equation}
one finds the microcanonical inverse temperature $\beta^*$.  For a finite 
system a plot of $E$ versus $1/\beta^*(E)$ (the microcanonical caloric curve)
shows a lot more detail than the 
equivalent canonical plot of $\left< E \right>$ as a function of $T$.
The sum in (\ref{Gl:3}) involves an average over many energy levels, whereas 
(\ref{Gl:9}) exhibits local anomalies of the density of states. There is
furthermore no need to determine the constants $K$ of MC runs 
at different temperatures. 
These drop out when the derivative is performed in (\ref{Gl:9}).
The bits of 
the $E$ versus $1/\beta^*$ curve from the different runs join smoothly 
without any further need for adjustment. 

Here we consider two and three dimensional Ising systems with 
ferromagnetic nearest neighbour interactions. The natural variables of the 
entropy are the energy $E$ and the magnetization $M$. We are interested
in the thermal behavior of a system with clamped magnetization $M = M_0$,
which can be deduced from $g(E,M_0)$, the degeneracy of the states 
with fixed magnetization as a function of the energy. 
Fixing the magnetization means fixing the number of minority spins.
At low energies almost all the minority spins form one big droplet
whereas at high energies one observes a large number of small droplets composed
of only a few minority spins.

In order to find $g(E,M_0)$ we perform a conventional canonical MC 
run at a temperature $T = 1/\beta$. At the start the system is
prepared in any one of the states with $M = M_0$ and with an  arbitrary energy 
$E_i$. During the procedure we randomly select one spin in the system and 
try to flip it. The flip leads to a new state with energy $E'$ and
magnetization $M'$. The flip is not performed if $M' > M_0 + 2$ or if
$M' < M_0 - 2$. If $M'$ is equal to one of the three admitted values 
$M_0$, $M_0 \pm 2$ of the
magnetization, the spin is flipped with probability 1 if $E' \leq E$,
and it is flipped with probability $\exp \left( \beta \left( E_i - E' \right)
\right)$ if $E' > E_i$. After every 1000 attempted
spin flips a histogram $h(E,M)$ is augmented by 1 count 
at the actual value of $E$ and $M$.
This way one obtains estimators for 
$g(E,M) \, \exp \left( - \beta \, E \right)$  for the three 
allowed values of the magnetization and 
for a range of energies centered around $\left< E \right>_\beta$
(\ref{Gl:7}). Alternatively, one might have used Kawasaki dynamics, but our
approach has the advantage that three independent histograms for the
three allowed values of the magnetization are obtained in one run.

The degeneracy $g(E,M_0)$ can now be exploited in different ways. One 
can e.g.\ calculate $\left< X \right>$ from (\ref{Gl:7}).
For large enough systems this 
corresponds to a sampling of $X(E)$ with a gaussian distribution 
function centered at the most probable value of $\tilde{E}$ of $E$
which solves $\mbox{d}S(E,M_0)/\mbox{d}E |_{\tilde{E}} = \beta$.
The width of the distribution
\begin{equation} \label{Glnew}
\Gamma_{can}=
\left( \mbox{d}^2 S(E,M_0)/\mbox{d}E^2 \right)^{-\frac{1}{2}} 
\end{equation} 
depends on the number $N$ of spins in the system, but it
tends to a constant for large system sizes.
On the extensive
energy scale $X(E)$ is therefore sampled over a constant range
$\Delta E$ of energies and the relative width of the distribution
$\Delta E/E_{max}$ with $E_{max}=d \, N$ ($d$ being the 
dimensionality of the lattice) tends to zero with increasing
system size.

One may also calculate local derivatives of $S(E,M_0)$ as e.g.\ the 
microcanonically defined temperature $\beta^*(E,M_0) = S'(E,M_0)
= \mbox{d}S(E,M_0)/\mbox{d}E$ or its derivative $\mbox{d} \beta^*(E,M_0)/
\mbox{d}E = S''(E,M_0)= \mbox{d}^2S(E,M_0)/\mbox{d}E^2$ 
which serve in the definition of the specific heat:
\begin{equation} \label{Gl:10}
C(E)= - S'(E,M_0)^2/S''(E,M_0).
\end{equation} 

Statistical errors in the numericaly determined histogram $g(E,M_0)$
are unavoidable. These are accentuated when derivatives are 
formed. The fluctuations of $S'(E,M_0)$ and $S''(E,M_0)$ can be 
smoothed by gaussian sampling:
\begin{equation} \label{Gl:11}
\left\{ S'(E,M_0)  \right\} = \sum\limits_i \, S'(E_i,M_0) \,
\exp \left( - \left(E_i - E \right)^2/(2 \Gamma^2 ) \right)/Q_m
\end{equation}
where
\begin{equation} \label{Gl:12}
Q_m = \sum\limits_i \exp \left( - \left( E_i - E \right)^2/(2 \Gamma^2 ) \right)
\end{equation}
is the normalization. Eqs.\ (\ref{Gl:11}) and (\ref{Gl:12}) 
bear some resemblence 
with the canonical mean values of (\ref{Gl:7}) and (\ref{Gl:8}),
but here we have 
the advantage that the center $E$ and the width $\Gamma$ of the gaussian 
distribution can be chosen at will - as long as the sampling is 
restricted to energy levels $E_i$ with sufficiently accurate data.

The freedom of choosing the width $\Gamma$ is exploited in a 
compromise between the demand of exhibiting the local anomalies of the 
entropy clearly and the goal of getting rid of the statistical errors as 
well as possible. Gaussian smoothing has the definite advantage over any 
other smoothing procedure that it does not create any spurious 
features which were not present in the original data. In the canonical
averaging procedure (\ref{Gl:7}) $\Gamma_{can}$ (see (\ref{Glnew})) is
fixed. The value of $\Gamma$ in (\ref{Gl:12}) should be as small as possible,
but this depends on the quality of the data. In the examples of the
next Section we were able to keep $\Gamma$ two orders of magnitude
smaller than $\Gamma_{can}$. Therefore much finer details of the
caloric curve can be resolved.

Besides using the procedure just described, we also simulated the IMFM model
with an adapted nonlocal spin-exchange algorithm\cite{Bar94}, especially designed for
the investigations of equilibrium crystal shapes, in order to study
the thermal stability of extensive macroscopic droplets in three dimensions.
 
\newpage

\section{The caloric curve}
Recently, the thermal stability of macroscopic droplets of minority spins
in small systems has been investigated at fixed magnetization in the two-dimensional
Ising model\cite{Ple00}.
Inside the coexistence region
the disappearance of
these droplets is accompanied by a pronounced maximum in the specific heat.
Similar peaks were also observed in three dimensions\cite{Car98}.
However, rigorous results\cite{Dob92,Bod00,Dob94,Iof98} show that for the infinite two-dimensional lattice
this transition takes place at the coexistence line. The same conclusion may be drawn
from a rather simple and physically appealing argument which we will briefly discuss
in the next section when dealing with the stability of three-dimensional
droplets. One may therefore ask
how to interpret these specific
heat peaks in the light of what is known about this line.

As a thorough investigation of the caloric curve should shed new light
on this problem, we studied the two- and the three-dimensional IMFM using the
histogram method presented in the previous Section. Two-dimensional square lattices 
with $N=L^2$ spins and three-dimensional simple cubic lattices with $N=L^3$ spins,
both with full periodic boundary conditions, were simulated. The system sizes
ranged from $L= 100$ to $L=900$ in two dimensions and from $L=10$ to $L=80$
in three dimensions.

Fig.\ 1 shows the computed microcanonical 
caloric curve $e$ versus $1/\beta^*$ for the two-dimensional
system at the fixed magnetization density $m_0=M_0/N=0.92$ for different system sizes. Here, $\beta^*$ is the 
microcanonical inverse temperature defined in eq.\ (\ref{Gl:9}), whereas
$e$ is the intensive energy $E/N$.
We emphasize again
that the Gaussian smoothing procedure (see Section 2) used in this plot 
only reduces the unavoidable statistical errors.
All the features discussed in the following are clearly visible in our original,
unsmoothed, high-quality data.

\begin{figure*}[h]
\centerline{\psfig{figure=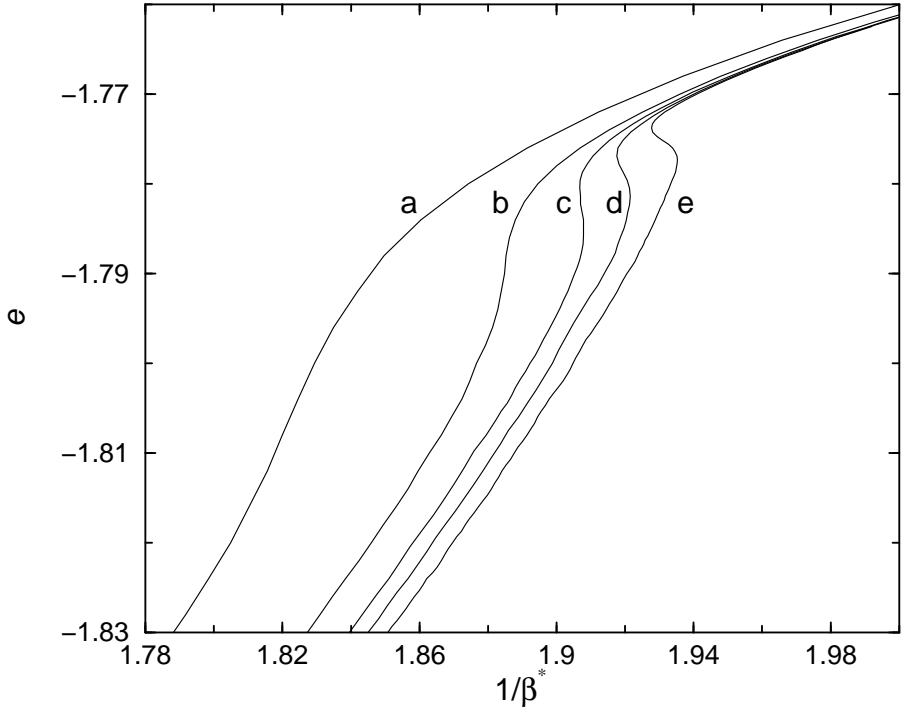,angle=270,width=12cm}}
\caption{Microcanonical caloric curves of the two-dimensional Ising model
at the fixed magnetization $m_0=0.92$ for different system sizes: ({\bf a}) $L=100$,
({\bf b}) $L=200$, ({\bf c}) $L=300$, ({\bf d}) $L=400$, and ({\bf e}) $L=600$.
}
\label{fig1} \end{figure*}

Inspection of Fig.\ 1 reveals a completely different behavior of the system for
small and large system sizes.
For not too large systems (line {\bf a} in Fig.\ 1)
the caloric curve is monovalued, showing no indications of a possible phase transition.
Increasing $L$ leads first to a steepening of the curve (line {\bf b}). After
a threshold value has been passed, 
the monovaluedness is lost and the characteristic S-shape of a 
first order phase transition in finite systems\cite{Hue94}
is observed (lines {\bf c}, {\bf d}, and {\bf e}). This 
backbending of the caloric curve is associated with a segment
of positive curvature of the entropy and leads to a negative microcanonical specific heat,
see eq.\ (\ref{Gl:10}). Remarkably,
large system sizes are needed in order to exhibit the S-shaped microcanonical
caloric curve as the threshold value is around $L=250$. 
Usually, in first order phase transitions, this characteristic behavior
is already encountered for a small number of spins\cite{Sch94}. 
It is worth mentioning that the canonical caloric curve does not show
this backbending for any system size\cite{Hue94}. One should further note that
the same systematics have been observed, as a function of the number of atoms, in the theoretical
study of the solid-liquid transition in Lennard-Jones clusters\cite{Lab90}.

The convex intruder, where the low- and the high-temperature phases
coexist, is shown in Fig.\ 2 for two different system sizes. Here, the quantity $N \left( s -
\beta_t e \right)$ is plotted as a function of the intensive energy.
$s$ is the microcanonical entropy $s(e, m_0) = N^{-1} \ln g(e, m_0)$ whereas
the transition temperature $\beta_t^{-1}$ is obtained by the double tangent
construction: On the major part of the energy axis $s(e)$ is a concave function
of $e$. There the tangents of $s(e)$ lie completely above $s(e)$, in the convex
part any tangent intersects $s(e)$. Between the two cases there is one tangent which
touches $s(e)$ at the two points with the energies $e_1$ and $e_2$. This is the double 
tangent which determines the transition temperature $\beta_t^{-1}$ and the finite
system latent heat\cite{Hue94,Gro} $\Delta e$ at the same time:
\begin{equation} \label{Gl:13}
\beta_t = \frac{s(e_2)-s(e_1)}{e_2 - e_1}
\end{equation} 
and
\begin{equation} \label{Gl:14}
\Delta e = e_2 - e_1.
\end{equation}
Note that the curves in Fig.\ 2 are raw, unsmoothed curves, demonstrating the good quality
of our data.

\begin{figure*}[t]
\centerline{\psfig{figure=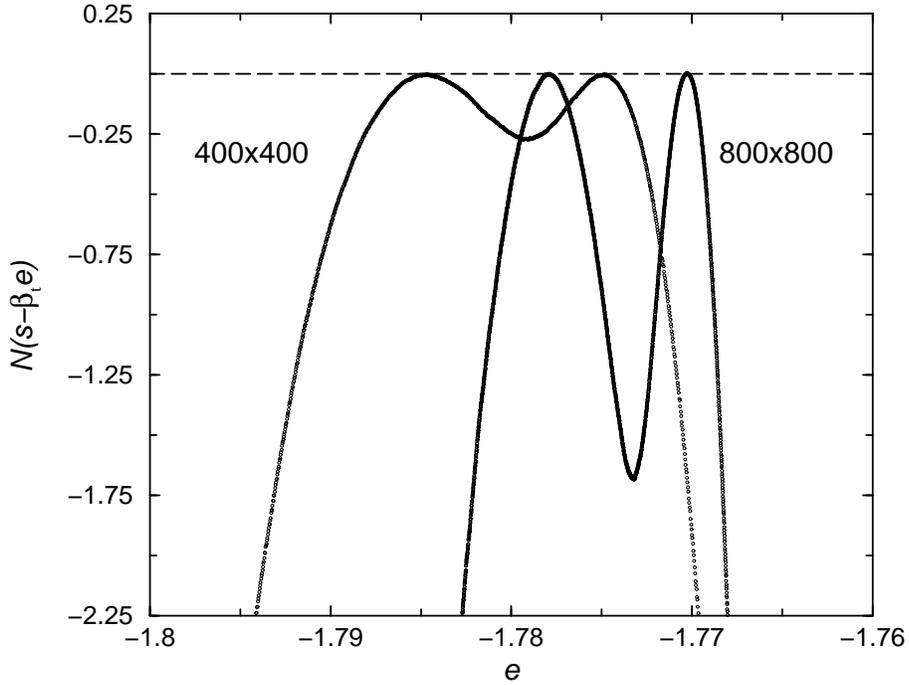,angle=270,width=12cm}}
\caption{The function $N(s - \beta_t e)$ versus the inverse energy $e$ for
the two-dimensional Ising model at constant magnetization $m_0=0.92$ for two
different system sizes. We show the unsmoothed data, thus demonstrating their
high quality.}
\label{fig2}  \end{figure*}

From the finite-size scaling theory for first order phase transitions\cite{Lee91,Bor90,Bor92,Bor91}
it follows that the energies $e_1$ and $e_2$ should tend
to their bulk values linearly in $L^{-1}$,
whereas their difference, i.e. the latent heat (see eq.\ (\ref{Gl:14}))
must approach a constant value if the transition is really
discontinuous. Furthermore, the asymptotic behavior of the transition
temperature $\beta_t^{-1}$ should be proportional to $L^{-2d}$ ($d$ being the 
dimensionality of the lattice) \cite{Bor90,Bor92}, whereas the depth of the minimum of the function
$s - \beta_t e$ should vanish as $L^{-1}$. In the present case, however,
the situation is more complicated: The asymptotic behavior seems to be
encountered only for the largest systems considered (difficulties in observing the
asymptotic behaviour in first order phase transitions have also been reported
in other systems\cite{Ste98}). As an example, Fig.\ 3
shows the evolution of the energies $e_1$ and $e_2$ as a function
of $L^{-1}$. For systems with less than $800^2$ spins, the energies $e_1$
and $e_2$ do not vary linearly in $L^{-1}$. The distance between these
two energies,  i.e. the latent heat, first increases, reaches a maximum for a system
with about $400^2$ spins, then decreases until, for $L=800$, a linear dependency seems to
prevail. Larger system sizes, which are beyond our present computer facilities,
are needed to affirm unambiguously that the asymptotic regime has been reached.

\begin{figure*}[t]
\centerline{\psfig{figure=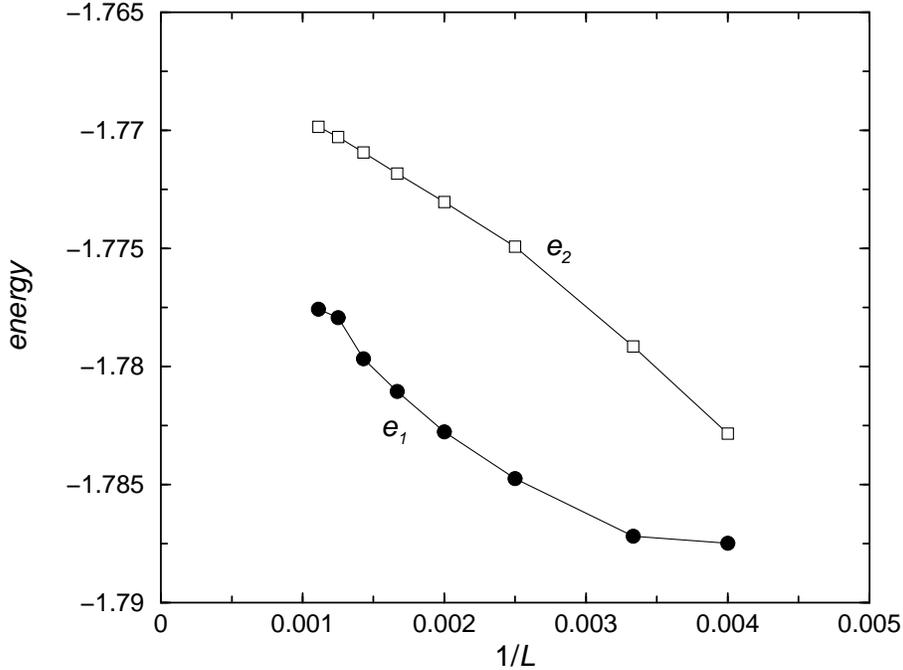,angle=270,width=12cm}}
\caption{Energies $e_1$ and $e_2$ as function of the inverse system length $L^{-1}$
for the two-dimensional Ising model with $m_0=0.92$. The difference $e_2 - e_1$
is the latent heat. Error bars are smaller than the symbol sizes.
}
\label{fig3} \end{figure*}

Turning now to the three-dimensional case, we show in Fig.\ 4 the microcanonical
caloric curve for the fixed magnetization density $m_0=0.568$ as a function of the linear size $L$
of the system. The same systematic behavior as in two dimensions is observed.
Around $50^3$ spins are needed in order to observe the typical S-shape of a continuous
transition. Fig.\ 5 shows the caloric curve of systems with $L=60$ and
different values of the fixed magnetization. At higher magnetizations, the phase
transition takes place at lower values of the microcanonical temperature $1/\beta^*$,
as would have been expected from the shape of the coexistence line. Remarkably,
the discontinuous character of the transition is, at a given system size,
much more pronounced at larger values of the magnetization. For small magnetization, see
line {\bf f} in Fig.\ 5, the caloric curve for $L=60$ is monovalued and does not present
the backbending. Clearly, larger system sizes have to be simulated when the magnetization
is decreased in order to observe the characteristics of a possible first order phase transition. 
This may be related to the fact that along the coexistence line 
the correlation length increases with decreasing magnetization, diverging finally
at the critical point. 

\begin{figure*}[t]
\centerline{\psfig{figure=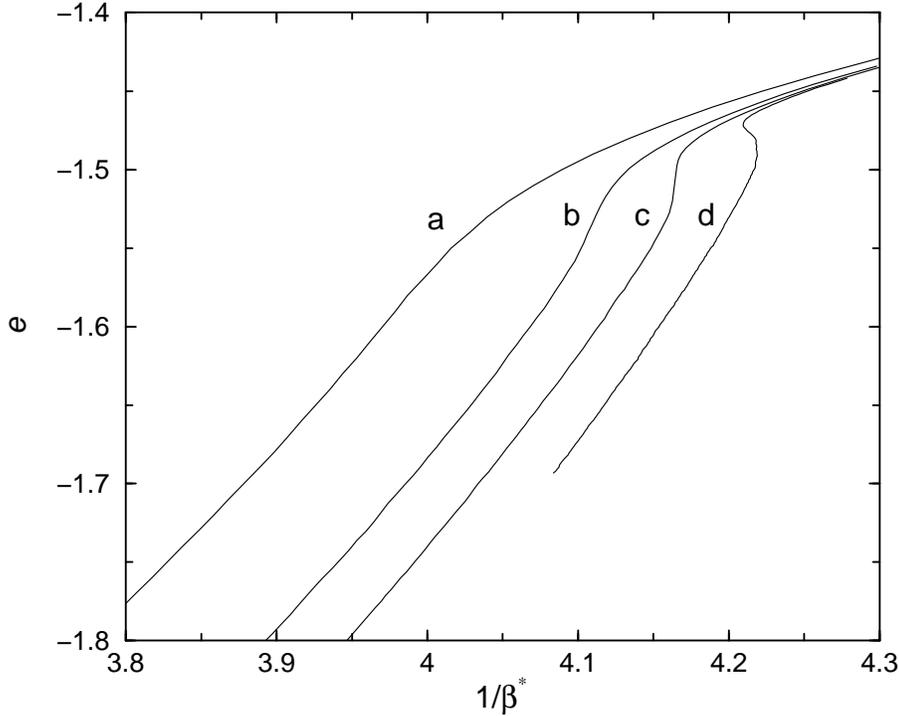,angle=270,width=12cm}}
\caption{Microcanonical caloric curves of the three-dimensional Ising model
at the constant magnetization $m_0=0.568$ for system sizes
({\bf a}) $L=20$, ({\bf b}) $L=30$, ({\bf c}) $L=40$, and ({\bf d}) $L=60$.
}
\label{fig4} \end{figure*}
 
\begin{figure*}[h]
\centerline{\psfig{figure=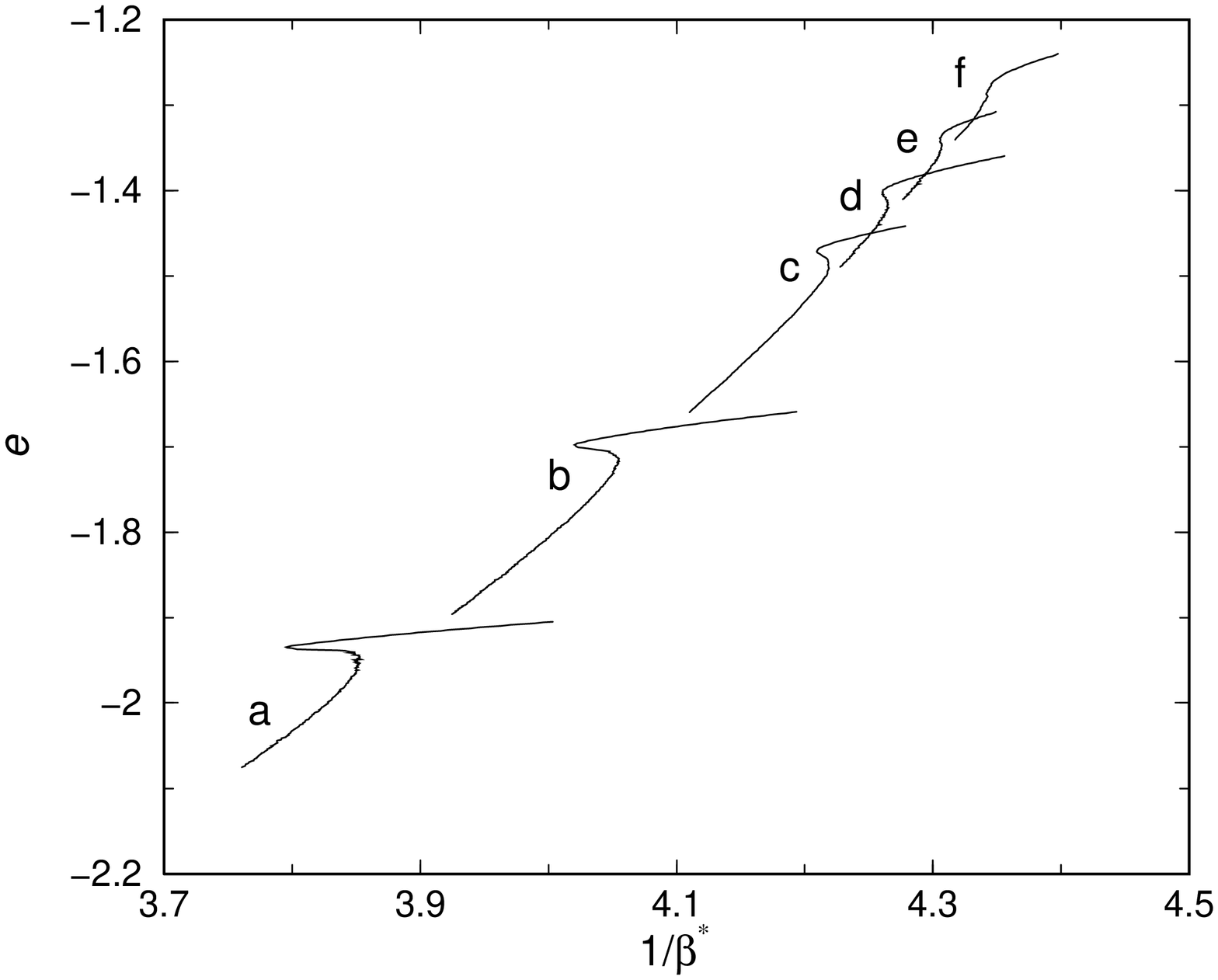,angle=270,width=12cm}}
\caption{Microcanonical caloric curves obtained for a system with $60^3$ spins
for various fixed magnetizations: ({\bf a}) $m_0=0.75$, ({\bf b}) $m_0=0.66725$,
({\bf c}) $m_0=0.568$, ({\bf d}) $m_0=0.531$, ({\bf e}) $m_0=0.492$, and
({\bf f}) $m_0=0.45075$.
}
\label{fig5} \end{figure*} 

Our data for large system sizes are in accordance with conclusions 
drawn from the expected topology of the
entropy in finite systems\cite{Gro00}.
With the two-phase region excepted, the entropy is a concave function of the
energy $e$ and the magnetization density $m$, i.e.\ both the principal
curvatures are negative. Inside the two-phase region of a finite system
one of the principal curvatures becomes positive.
Inspection of the eigenvectors
of the curvature matrix reveals that the convex intruder should then be encountered
when crossing the coexistence line at fixed magnetization, as is observed
in our simulations of large systems. On the other hand, our simulations
of small systems show that the backbending
of the caloric curve is not visible for system sizes $L$ smaller than some 
characteristic, magnetization dependent, length which may be related to the 
correlation length. Thus,
our results explain why data from simulations of three-dimensional
moderate sized systems (up to $40^3$ spins)
at low magnetizations were erronously 
interpreted as evidence for a continuous phase transition\cite{Car98}
and they refute the recent claim 
that at constant volume in the corresponding lattice gas
model the microcanonical caloric curve does not possess the typical S-shape\cite{Gul00}.

\section{Droplets in three dimensions}
In this Section, we will discuss the influence of finite-size effects on the
thermal stability, at fixed magnetization, of three-dimensional macroscopic droplets of minority spins 
inside the coexistence region of the Ising model\cite{Bod99,Cer00}. To be more specific, we will assume
a positive fixed magnetization $m_0$, so that the $\sigma=-1$ spins are the minority spins.
By the term ''macroscopic droplet'' we denote an extensive cluster of connected minority
spins with an average of $\left< N_c \right>$ spins 
whose size scales with the system size $N$. A corresponding study 
in two dimensions\cite{Ple00} has recently shown that these droplets loose their extensivity 
at a cluster transition which is expected to take place at the coexistence line
in the thermodynamic limit. In fact, one may argue that, in the infinite system, the 
mean magnetization density in the droplet
has the same absolute value as in the rest of the
system. As a consequence, the following expression for
the reduced droplet size $n_c = \left< N_c \right>/N_-$ ($N_-$ being the total number of
$\sigma=-1$ spins) in the thermodynamic limit
is easily derived\cite{Ple00}:
\begin{equation} \label{Gl:15}
n_c(T)=(1+m_{sp})(m_{sp}-m_0)/(2m_{sp}(1-m_0))
\end{equation}
where $m_{sp}$ is the spontaneous magnetization at the temperature $T$. It follows
from this expression, in accordance with rigorous results,
that the extensive droplet vanishes in the infinite system at the coexistence line.
Note that no assumption on the dimensionality of the system was made for the derivation of
eq.\ (\ref{Gl:15}). In finite systems, this cluster transition is accompanied by
a pronounced peak in the specific heat (as obtained from energy fluctuations),
thus showing its connection to the transition discussed in the previous Section.

The assumption of equal absolute values of the magnetizations
inside and outside the droplet
is not expected to hold at high temperatures in finite systems. In fact,
due to the Gibbs-Thomson effect, which describes the effect of curved interfaces at
equilibrium\cite{Bin80,Kri96}, the fraction of minority spins forming the droplet at fixed 
magnetization $m_0$ should be reduced
in finite systems as compared to the infinite system, thus leading to a decrease 
of the reduced droplet size $n_c$ as well as to a shift of the cluster
transition temperature to lower values. This is exactly what has been observed
in simulations in two dimensions\cite{Ple00}.

Fig.\ 6 and Fig.\ 7 show the temperature dependent droplet size $n_c(L)$ for different
system sizes $L$ in the three dimensional case for two different fixed magnetizations
$m_0=0.75$ and $m_0=0.568$. This data were obtained with a nonlocal spin-exchange
algorithm\cite{Bar94}, specially conceived for the study of equilibrium crystal
shapes, which we adapted to the three-dimensional IMFM. The dashed lines 
are obtained from eq.\ (\ref{Gl:15}). 
For the spontaneous magnetization $m_{sp}(T)$ we used the expression given in Ref.\cite{Tal96}.

\begin{figure*}[!h]
\centerline{\psfig{figure=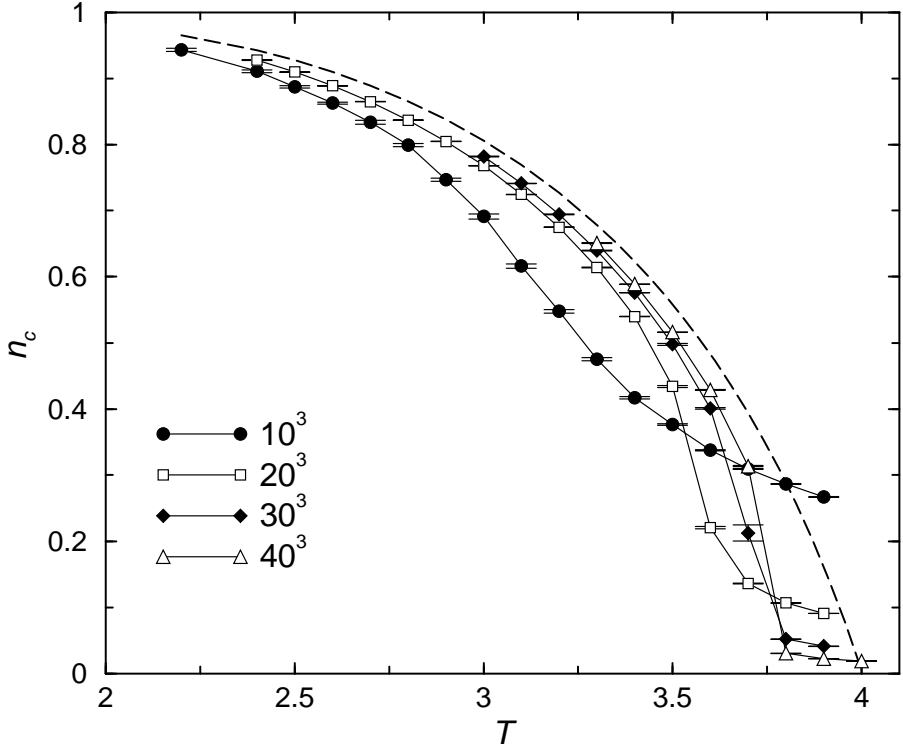,angle=270,width=12cm}}
\caption{Reduced droplet size $n_c$ as function of temperature for different
lattice sizes as obtained from Monte Carlo simulations of the three-dimensional
IMFM with $m_0=0.75$. Here and in the following error bars are obtained by
averaging over at least 10 runs with different random numbers. The dashed
line is the theoretical prediction (\ref{Gl:15}) for the infinite system.
}
\label{fig6} \end{figure*}
 
\begin{figure*}[!h]
\centerline{\psfig{figure=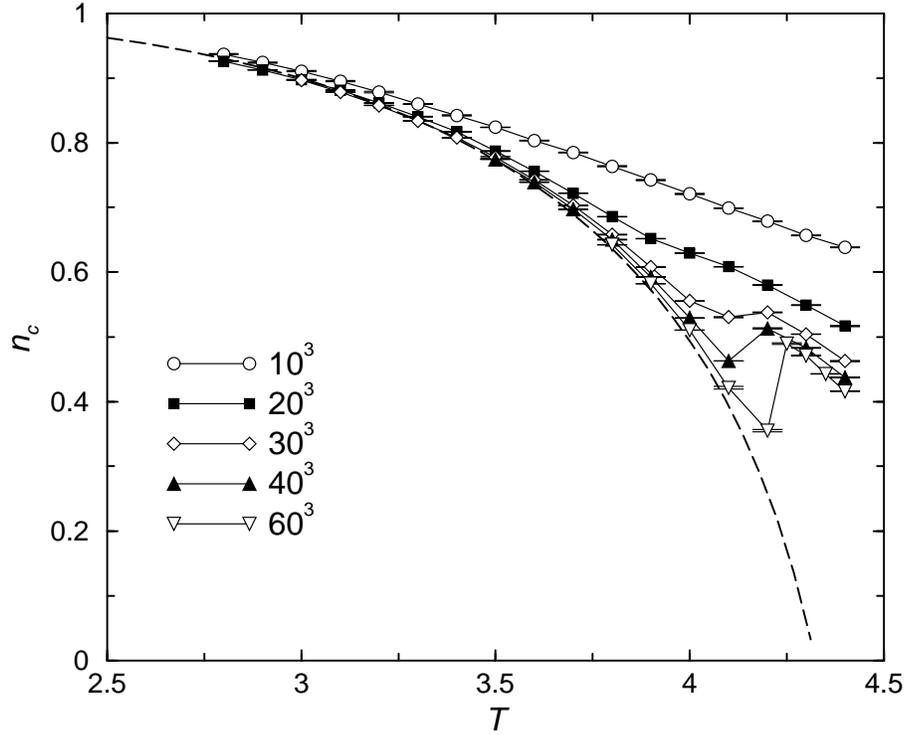,angle=270,width=12cm}}
\caption{The same as in Fig.\ 6, but now for the conserved magnetization $m_0=0.568$.
}
\label{fig7} \end{figure*}

Consider first the case $m_0=0.75$ (Fig.\ 6). For this value of the fixed magnetization,
the reduced droplet size $n_c(L)$ for finite $L$ is shifted to lower
values when compared to the theoretical
prediction (eq.\ (\ref{Gl:15})), but it clearly approaches the theoretical curve when
$L$ is increased. Furthermore, the cluster transition temperature, given by the turning
point of $n_c(L)$, is reduced for decreasing system sizes. These results
are readily understood when invoking the Gibbs-Thomson effect and are similar to the 
finite size dependences observed in two dimensions. Note that for increasing system
sizes, the decrease in $n_c(L)$ evolves from a slow smooth decrease for small systems to a rapid, 
almost discontinuous, change for larger systems, in accordance with the systematics discussed
in the previous Section. 			

It has been proven rigorously\cite{Dob92,Dob94} that at fixed magnetization density $m_0$
in a $d$-dimensional hypercube with $L^d$ sites ($L$ very large, $L \longrightarrow 
\infty$) one big droplet is observed as long as
\begin{equation}
m_{sp}(T) - m_0 \gg L^{-\frac{d}{d+1}}
\end{equation}
whereas many small droplets with sizes at most as large as $\ln L$ are encountered if
\begin{equation}
m_{sp}(T) - m_0 \ll L^{-\frac{d}{d+1}}.
\end{equation}
Setting $m_{sp}(T) - m_0 = L^{-\frac{d}{d+1}}$ one may derive for a given $m_0$ and a given $L$
a typical temperature separating these two regimes. These temperatures
coincide rather well with the transition temperatures defined as the turning points
of the reduced droplet sizes, see Fig.\ 6, even if the rigorous proof only holds in
the limit $L \longrightarrow \infty$. The same is true for droplets in the two-dimensional
Ising model, see Fig.\ 2 in Ref.\ \cite{Ple00}.

For $m_0=0.568$ (Fig.\ 7), the finite system sizes manifest themselves in a completely different
and unexpected way. The reduced droplet size now approaches the theoretical curve from above,
i.e. the fraction of minus spins forming the droplet is larger in finite systems than compared
to what is expected from the assumptions leading to eq.\ (\ref{Gl:15}). This behavior 
is in marked contrast to the Gibbs-Thomson effect.

Our explanation of this behavior is based on percolation theory\cite{Sta94}.
Recall that in the three-dimensional Ising model, the percolation temperature $T_p$ for the
type of clusters we are considering (i.e.\ clusters of minority spins connected by a 
nearest-neighbor bond) is located below the critical temperature $T_c$\cite{Mue74,Con77}. Therefore, for temperatures
larger than $T_p$, an infinite cluster of minority spins,
which is not a compact object but a fractal, is encountered at the coexistence line
in the infinite system. In the present case, the crossing of the line $m_0 = 0.568$ and the
coexistence line is located slightly above the percolation temperature. It is then not surprising
to observe that the extensive droplet contains a larger fraction of minority spins than predicted
under the assumptions (involving implicitly compact objects) leading to eq.\ (\ref{Gl:15}).
In fact, above the percolation temperature, eq.\ (\ref{Gl:15}) should not be valid any more.
The data in Fig.\ 7 show that in finite systems even at temperatures below $T_p$ the droplet has a tendency
to be formed by an increased number of minority spins. These conclusions are supported by results
for the reduced size of the largest cluster formed by minus spins in the two-dimensional IMFM at
magnetization $m_0=0$. 
Note that for the two-dimensional Ising model the critical temperature and the
percolation temperature coincide. In the simulations of the two-dimensional IMFM with $m_0=0$ 
larger clusters than theoretically predicted are also observed, in accordance
with the fact that the $m_0=0$ line crosses the line of spontaneous magnetization at the
critical (percolation) temperature, whereas
for all other investigated values of the fixed magnetizations $m_0 \neq 0$ a decrease
of the reduced droplet size is encountered when compared to the theoretical 
line (\ref{Gl:15}) \cite{Ple00}.

Coming back to Fig.\ 7, we observe further for large system sizes a discontinuous jump in $n_c(L)$. This
jump is related to the transition encountered when the coexistence line is crossed,
as discussed in Section 3.
In contrast to the case of larger values of $m_0$ (see Fig.\ 6), where the 
discontinuous change is at best guessed for large system sizes, the discontinuity in the droplet size
is here clearly observed. This is a consequence of the largeness of the droplets of minority spins when crossing
the coexistence line in the vicinity of the percolation temperature. Note that
discontinuous changes are also observed in the second moments of the cluster
size distributions.

\section{Conclusions}
Despite of the fact that the Ising model has been studied extensively
since more than 70 years and despite of its possible applications in
solid state and nuclear physics, there are still some open questions regarding its
behaviour at constant magnetization $m_0$ in finite systems. In three dimensions it may
serve as a model system for the solid to gas transition of a fixed
number of particles in a given volume or for the fragmentation of
nuclear matter. The two dimensional Ising model is used to describe the
transition from a cluster of adatoms on a surface to a gaslike
behaviour.

We have performed a series of MC calculations in two and three
dimensions for several values of $m_0$ and for a wide range of system
sizes. For large systems the data hint at a possible discontinuous phase
transition with a well resolved latent heat. The discontinuity
disappears for smaller systems and it diminishes when $m_0$ is reduced.

It is well known that the canonical ensemble is ill prepared for a
distinction between continuous and discontinuous phase transitions\cite{Bin92}.
All singularities are smeared out. It is therefore essential
that the data, although obtained in normal canonical simulations, be
analyzed microcanonically.

There a possible discontinuous transition is clearly visible in finite systems as a convex
intruder in the entropy as a function of the energy (see Fig.\ 2). In a
canonical analysis one averages over a great number of energy channels
leading to a loss of information which is especially bad in the two
phase region between the two maxima of $s(e) - \beta_t e$. In the vicinity
of the transition temperature $T_t = 1/\beta_t$ the partition function is
dominiated by one or the other of the two maxima.

When the energy is plotted versus the microcanonically defined temperature,
the convex intruder manifests itself as an S-shape in the caloric
equation. This is shown in Figs.\ 1, 4, and 5.

There is no discontinuity present for small systems -- it
develops only for astonishingly large systems. Despite of the very large
system sizes which have been studied, we are not able to assert that the
scaling region has already been reached, but larger systems surpass our
present computer capacities. Fig.\ 5 shows that, for a given size, the
discontinuous character diminishes when $m_0$ is reduced. On the basis of
the present data we cannot decide if for very large systems there is a
discontinuity for all finite $m_0$ or if it disappears before the limit
$m_0 = 0$ is reached. For sufficiently large systems and for sufficiently
large values of $m_0$ a discontinuity is observed both in two and
three dimensions.

These results also explain why the decay of the macroscopic droplets of minority spins
with rising energy inside the coexistence region of finite systems is signalled by a peak in the
specific heat. In fact, these droplets loose their extensivity at the coexistence
line when crossing this line at temperatures below the percolation temperature. Above the percolation
temperature, the macroscopic Ising clusters percolate the system, the 
crossing of the coexistence line showing up as discontinuous jumps in the
cluster properties.

\section*{Acknowledgement}
We would like to thank D.\ H.\ E.\ Gross, M.\ Henkel, H.\ M\"{u}ller-Krumbhaar, J.\ Richert,
W.\ Selke, and Y.\ Velenik for useful discussions and D. d'Enterria for bringing Ref.\cite{Ric00} to our
attention.


\end{document}